\documentclass[preprint,aps,tightenlines]{revtex4}
\usepackage{amsmath}
\usepackage{verbatim}
\usepackage{graphicx}
\usepackage[usenames]{color}
\usepackage{psfrag}
\usepackage{graphicx}
\usepackage{natbib}
\usepackage{hyperref}
\usepackage{epstopdf}

\newcommand{\beq}{\begin{equation}}
\newcommand{\eeq}{\end{equation}}

\begin{document}

\title{Induced CMB quadrupole from pointing offsets}

\author{Adam Moss} \email{adammoss@phas.ubc.ca}
\affiliation{Department of Physics \& Astronomy\\
University of British Columbia,
Vancouver, BC, V6T 1Z1  Canada}

\author{Douglas Scott}  \email{dscott@phas.ubc.ca}
\affiliation{Department of Physics \& Astronomy\\
University of British Columbia,
Vancouver, BC, V6T 1Z1  Canada}

\author{Kris Sigurdson}  \email{krs@phas.ubc.ca}
\affiliation{Department of Physics \& Astronomy\\
University of British Columbia,
Vancouver, BC, V6T 1Z1  Canada}

\date{\today}

\begin{abstract}
Recent claims in the literature have suggested that the {\it WMAP\/} quadrupole 
is not primordial in origin, and arises from an aliasing of the much 
larger dipole field because of incorrect satellite pointing. We attempt to 
reproduce this result and delineate the key physics leading to the 
effect. We find that, even if real, the induced quadrupole would be smaller
than claimed.  We discuss reasons why the {\it WMAP\/} data are unlikely to
suffer from this particular systematic effect, including
the implications for observations of point sources.  Given this evidence
against the reality of the effect,
the similarity between the pointing-offset-induced signal and 
the actual quadrupole then appears to be quite puzzling. However, we find 
that the effect arises from a convolution between the gradient of the 
dipole field and anisotropic coverage of the scan direction at each 
pixel. There is something of a directional conspiracy here -- the dipole 
signal lies close to the Ecliptic Plane, and its direction, together 
with the {\it WMAP\/} scan strategy, results in a strong coupling to the 
$Y_{2,\,-1}$ component in Ecliptic co-ordinates. 
The dominant strength of this component in the measured quadrupole suggests 
that one should exercise increased caution in interpreting its 
estimated amplitude.  The {\it Planck\/} satellite has a different scan
strategy which does not so directly couple the dipole and quadrupole in
this way and will soon provide an independent measurement.
\end{abstract}

\maketitle
 
 \section{Introduction}
 
Measurements of Cosmic Microwave Background (CMB)
anisotropies from space have laid the foundations 
of the standard model of cosmology. These observations provide prima 
facie evidence that the Universe is close to spatially flat with nearly 
scale-invariant initial density fluctuations. It is remarkable that 
only a handful of other parameters, specifying the fraction of baryonic
and dark matter components, together with the local expansion rate and 
the redshift of reionization, can fit the data over such a wide range 
of scales. 
 
Despite the success of the standard model (see e.g~\cite{scott}), 
cosmologists have searched for evidence of new physics. Unfortunately 
this practice is fraught with uncertainty because of the prevalence of 
a posteriori statistics. One particular avenue of investigation has 
focused on the CMB anisotropies at large angular scales. A low 
quadrupole, for example,  was first noted by the Cosmic Background 
Explorer ({\it COBE})~\cite{Smoot:1992td} and subsequently confirmed by the 
Wilkinson Microwave Anisotropy Probe ({\it WMAP})~\cite{Bennett:2003ba}. 
Several other features have been discovered in {\it WMAP\/}
data (see~\cite{Bennett:2010jb} and references therein), with
subsequent debate about their statistical significance. 
Hence it is crucial to investigate all potential sources of systematic 
error which could be important at large angular scales. 
 
Recently, it has been claimed that an important effect has been 
overlooked in the {\it WMAP\/} analysis~\cite{Liu}. This particular issue 
relates to a 25.6\,ms offset between recording the pointing and 
differential temperature data from the satellite, which translates into 
an angular error of about $7^\prime$.
As part of the {\it WMAP\/} processing pipeline 
a dipole signal is first removed from the time-ordered data (TOD), 
and this has a much higher amplitude than the primordial anisotropies. If 
the pointing is incorrect a residual signal will remain in the TOD, 
which has been found to induce a quadrupole pattern in the final 
temperature sky maps. Interestingly, this quadrupole has  similar 
$a_{2m}$ spherical harmonic coefficients to those of
the {\em primordial} signal measured by WMAP. 

In this short article we show why the claimed result cannot be correct. 
In doing so we uncover the physics behind the coupling of such 
systematic effects to certain harmonic modes. We also investigate
the implications for the {\em Planck\/} experiment, which measures 
absolute rather than differential data, and scans the sky very 
differently to {\it WMAP}. 
 
 \section{WMAP results}
 
The {\it WMAP\/} TOD vector ${\bf d}$ can be written as 
\beq
{\bf d} = {\bf M} {\bf t} + {\bf n}\,,
\eeq
where ${\bf M}$ is the mapping matrix, ${\bf t}$ the sky-map and ${\bf 
n}$ the radiometer noise~\cite{Hinshaw:2003fc}. Ignoring the noise, the 
outputs of the two radiometers are~\cite{Jarosik:2007}
\beq
{\bf d}=\left( 1 + x_{{\rm im}} \right) {\bf i} (\hat{\bf n}_{\rm A}) - 
\left( 1 - x_{{\rm im}} \right) {\bf i} (\hat{\bf n}_{\rm B})\,, 
\eeq
where ${\bf i}$ is the Stokes $I$ parameter and the pointing on the sky 
is denoted by ${\hat{\bf n}}$, with the subscripts indicating the A and 
B-side beams. The pre-factor $(1 \pm x_{\rm im})$ for each term is due 
to transmission imbalance between radiometers,  the latest values for 
which can be found in~\cite{Jarosik:2010iu}. These factors are constant 
in time, and are easily absorbed into the mapping matrix.

For the purposes of this work we assume the noise is zero (including 
noise would only effect the statistics of the results, not the general 
features). The maximum likelihood sky-map solution is then $\tilde{\bf 
t} = \Sigma  \, \tilde{\bf t}_0$, where $\Sigma = \left( {\bf M}^{\rm 
T} \, {\bf M} \right)^{-1}$ and the `iteration-0' map $\tilde{\bf t}_0 
= {\bf M}^{\rm T} \,   {\bf d}$. The mapping matrix ${\bf M}$ has 
$N_{\rm t}$ rows and $N_{\rm p}$ columns, where $N_{\rm t}$ is the 
number of time samples for each radiometer and $N_{\rm p}$ the number 
of pixels in the map. 

To study the effects of a pointing offset, we assume our TOD contains a 
pure dipole signal. We fix the barycentric component to an amplitude 
3.3463\,mK with Galactic co-ordinates $(l, b)= (263.87^{\circ}, 
48.2^{\circ})$~\cite{Bennett:2003}. With a CMB monopole temperature of 
$T_{\rm 0}=2.725$\,K, this corresponds to a Galactic velocity ${\bf 
v}_{\rm gal} = \left( -26.2, 244.0, 274.4 \right)\,{\rm km}\,{\rm 
s}^{-1}$. We also include an additional varying dipole component due 
the satellite motion around the Sun, such that the total velocity is 
${\bf v} = {\bf v}_{\rm gal}  + {\bf v}_{\rm sol}$. The dipole at each 
pointing is then $(T_0/c) \,{\bf v} \cdot  {\hat{\bf n}}$ in 
temperature units.

From this simulated time-stream we then remove the dipole evaluated at 
a small pointing offset (corresponding to the claimed timing 
discrepancy), leaving a residual signal in the TOD.  We next run our 
map-making code, solving for the sky-map by a conjugate gradient 
method. We produce maps at a HEALPIX~\cite{Gorski:2004by} resolution of 
$nside=16$.
 
  \begin{figure}
\centering
\mbox{\resizebox{0.45\textwidth}{!}{\includegraphics[angle=90]{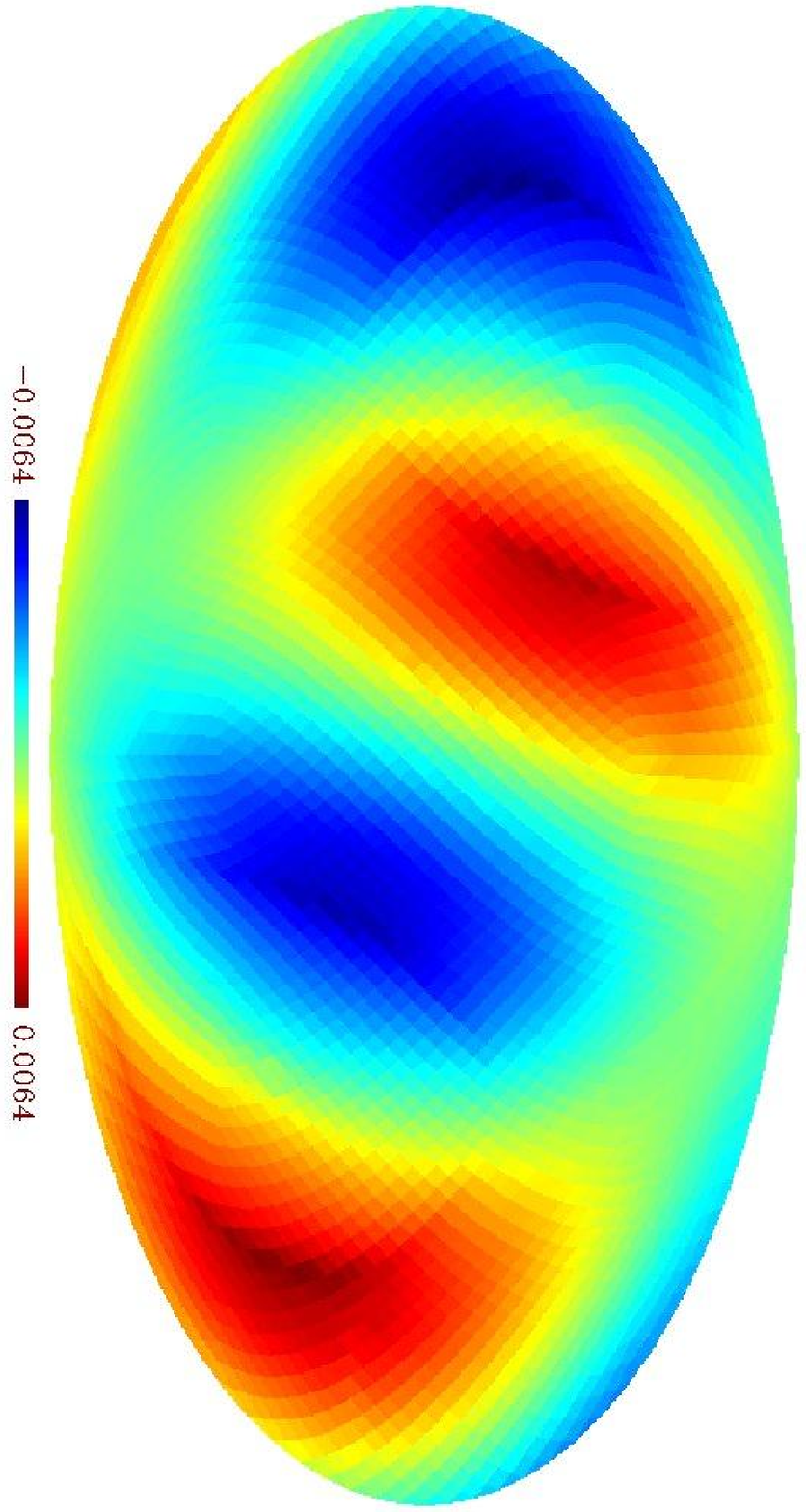}}}
\mbox{\resizebox{0.45\textwidth}{!}{\includegraphics[angle=90]{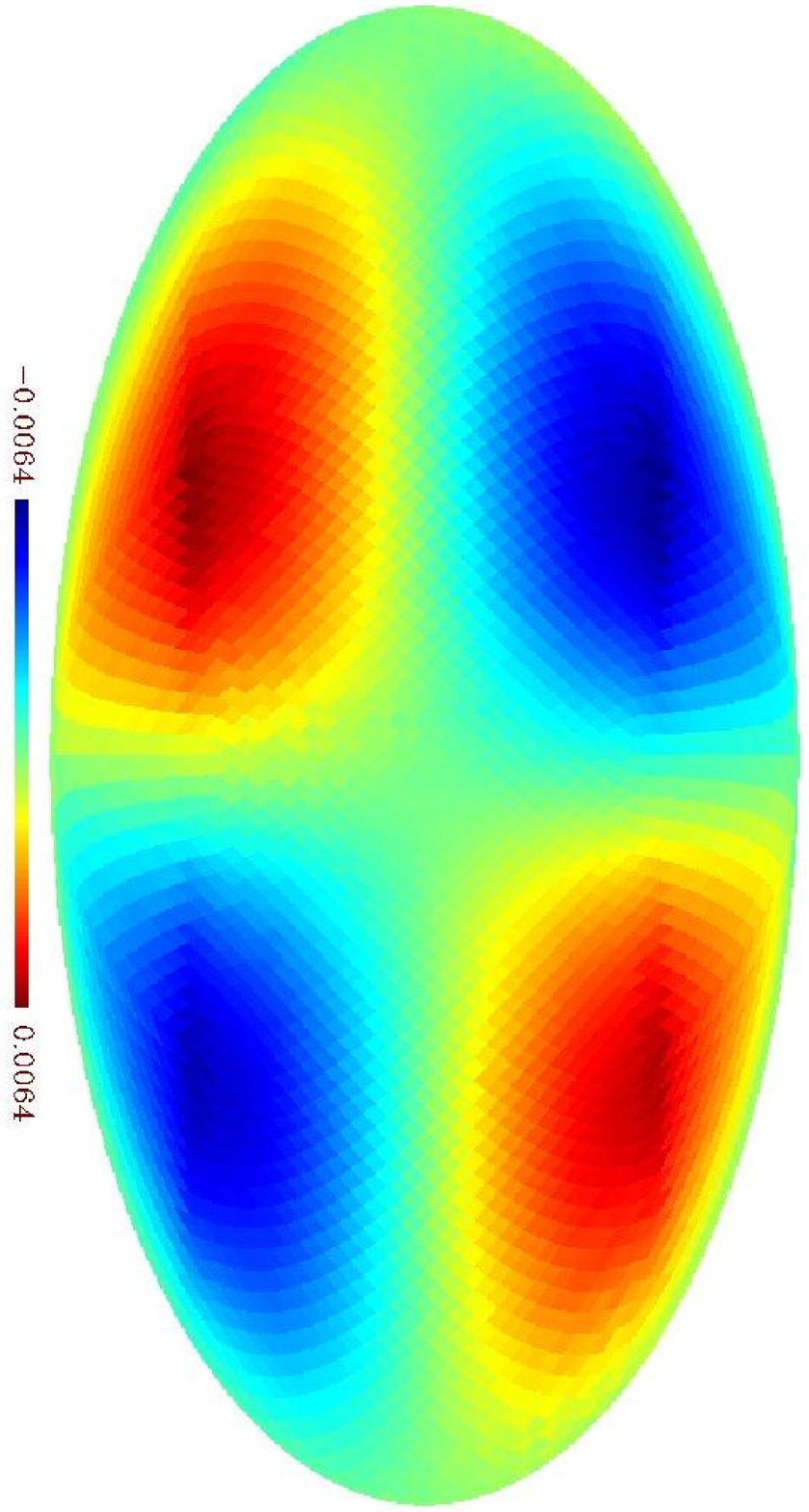}}}
\mbox{\resizebox{0.45\textwidth}{!}{\includegraphics[angle=90]{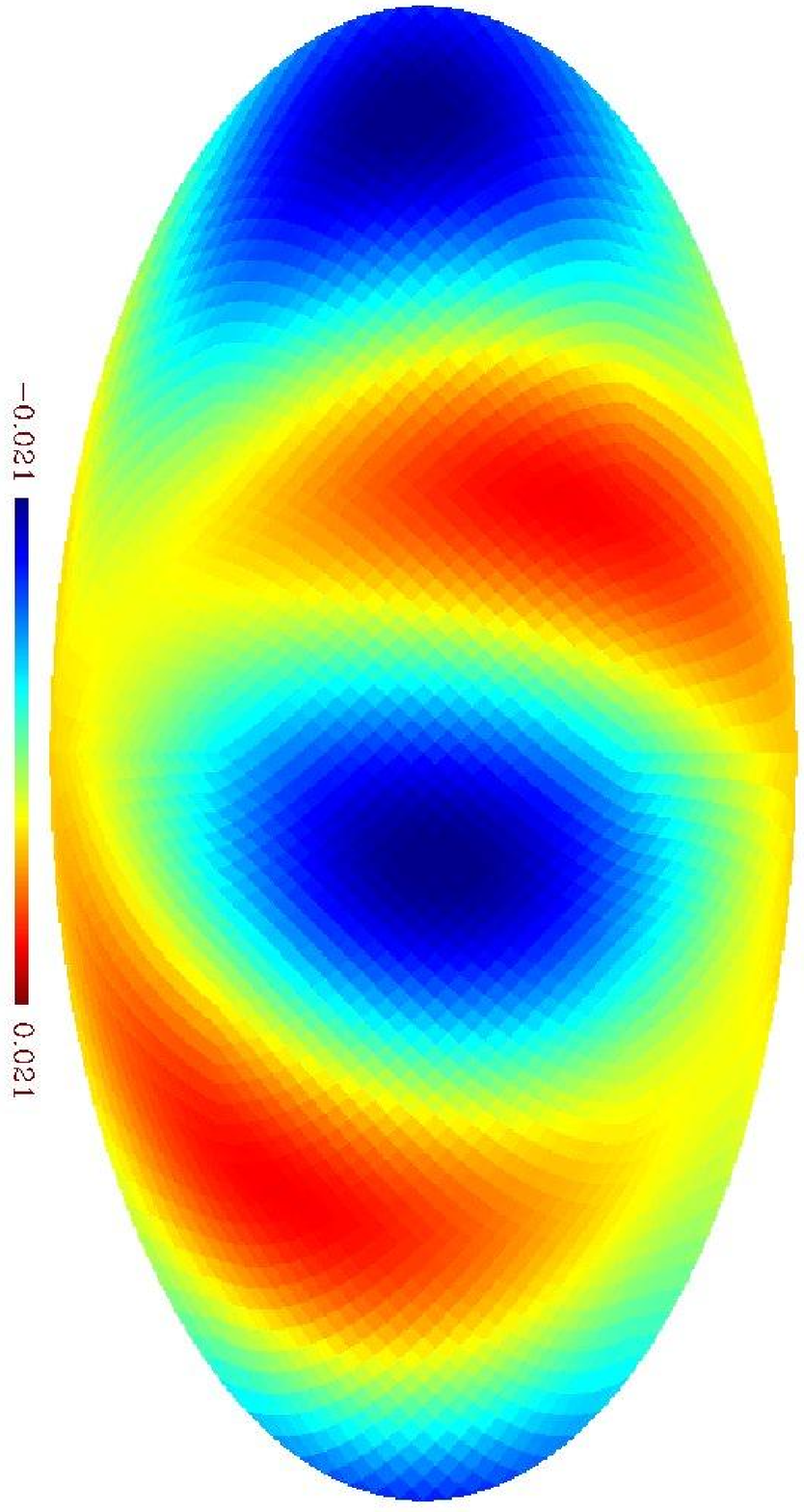}}}
\mbox{\resizebox{0.45\textwidth}{!}{\includegraphics[angle=90]{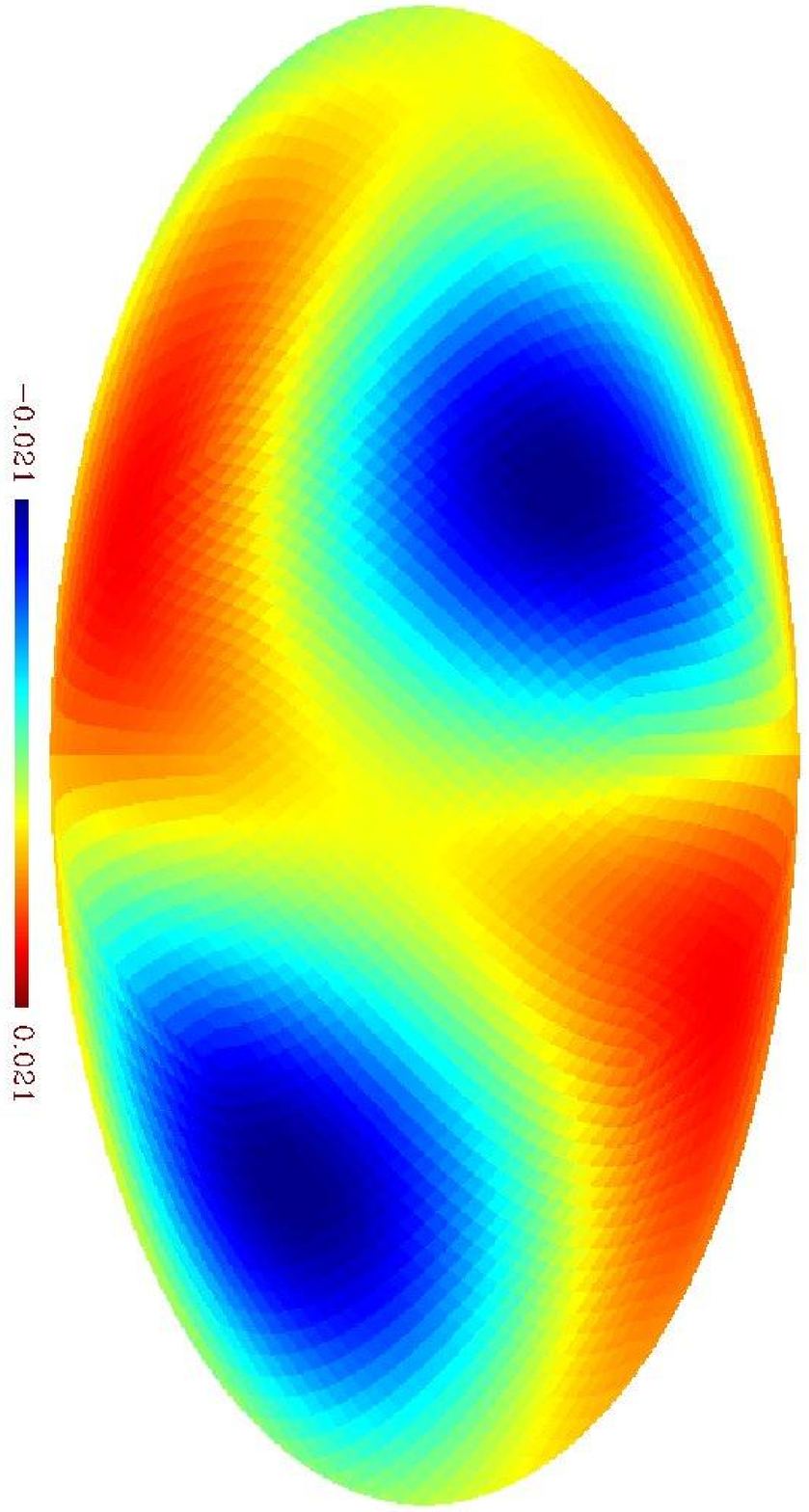}}}
\caption{\label{fig:W_solar} 
(Top): Induced quadrupole from a 25.6\,ms 
pointing offset in Galactic (left) and Ecliptic (right) co-ordinates. 
(Bottom): {\it WMAP\/} quadrupole derived from the 7-year ILC map in Galactic 
(left) and Ecliptic (right) co-ordinates. Units are in mK for each map.
The bottom panels are strongly correlated with the upper ones, although
with an amplitude about 3 times higher.\
 }
\end{figure}

Maps at W-band, which contain 30 observations (each lasting 51.2\,ms) 
per 1.536 second `science frame', are shown in Fig.~\ref{fig:W_solar}. In 
this case the central point of each observation corresponds to the 
claimed offset between recording the pointing information and recording the 
differential data from the satellite~\cite{Liu}. For W-band we can 
therefore use available {\it WMAP\/} software~\footnote{Available at 
http://lambda.gsfc.nasa.gov/product/map/current/m\_sw.cfm}~to compute 
the pointing offset, as the code also gives the interpolated pointing 
at the centre of each observation. The resulting map is characterized 
by a quadrupole with an amplitude $\sim 6.5 \,\mu$K, and in Ecliptic 
coordinates is close to a pure $Y_{2,\,-1}$ spherical harmonic mode. 
For comparison, we also plot in Fig.~\ref{fig:W_solar} (bottom panels)
the {\it WMAP\/} quadrupole derived from the 
7-year internal linear combination (ILC) map. 

The induced quadrupole maps are similar in structure to the published 
{\it WMAP\/} quadrupole. However, we note that we obtain an amplitude lower 
than that claimed in~\cite{Liu}, by approximately a factor of 3.
Hence, even if the claimed effect was 
real it could not have a large enough amplitude to explain the {\it WMAP\/}
result. Nevertheless, the fact that it would produce a quadrupole so 
similar is intriguing and seems worthy of further investigation.

\section{Understanding the induced quadrupole}

To make sense of this effect we built a toy model of the scanning strategy 
for satellites such as {\it WMAP\/} and {\em Planck}. The pointing vector can 
be obtained as a function of time by orienting the satellite focal 
plane (FP) in the $+z_{\rm FP}$ direction. The beam (or pair of beams 
in the case of {\it WMAP}) is specified by the opening angle $\theta_{\rm b}$ 
with respect to the $+z_{\rm FP}$ direction. In this frame the 
satellite spins at an angular velocity $\omega_{\rm s} = 2\pi/t_{\rm 
s}$, where $t_{\rm s}$ is the spin period. 
 
In order to achieve coverage at the Ecliptic poles the focal plane is 
rotated at a precession angle $\theta_{\rm p}$ with respect to the 
anti-solar direction (which is now defined as the new spacecraft (SC) 
$+z_{\rm SC}$ direction). The satellite then spins at a rate $w_{\rm 
p}=2\pi/t_{\rm p}$ in this frame. Finally, these co-ordinates are 
rotated into the Ecliptic Plane (with $+z$ corresponding to the north 
Ecliptic pole) and the solar rotation rate is fixed at $w_{\rm r}$. 

This toy model accurately reproduces the pointing of {\it WMAP\/} and
{\em Planck}. For {\it WMAP} \cite{Bennett:2003ba},
the relevant parameters are $\theta_{\rm 
b}=70^{\circ}$, $\theta_{\rm p}=22.5^{\circ}$, $t_{\rm s} = 129\,$s, 
$t_{\rm p} = 3600\,$s, while for {\em Planck\/} \cite{planck:scan}
they are $\theta_{\rm 
b}=85^{\circ}$, $\theta_{\rm p}=7.5^{\circ}$, $t_{\rm s} = 60\,$s and 
$t_{\rm p} = 6\,$months.  Note that the precession period is much 
longer for {\em Planck\/} than {\it WMAP}. 

\begin{figure}
\centering
\mbox{\resizebox{0.45\textwidth}{!}{\includegraphics[angle=90]
{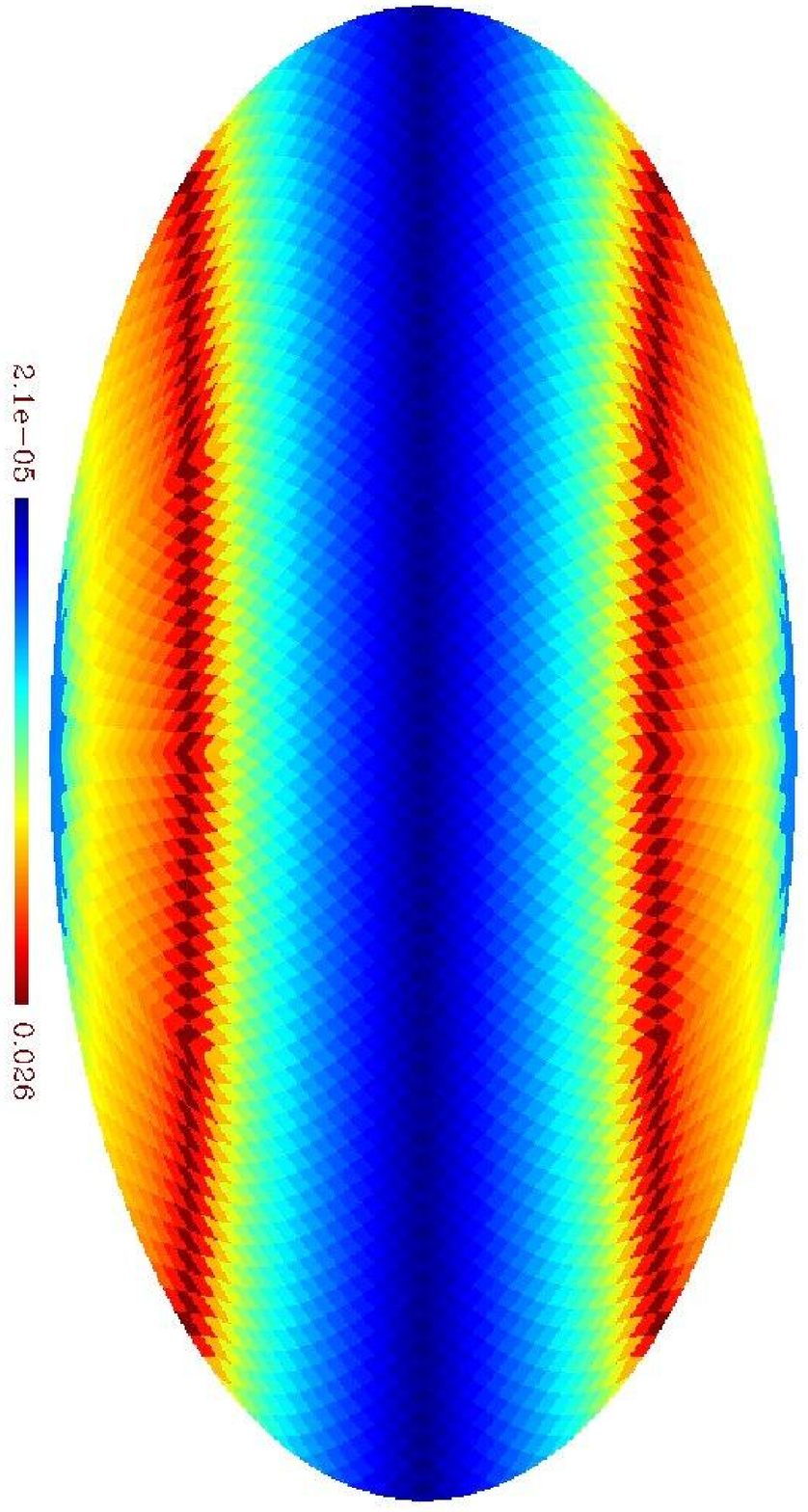}}}
\mbox{\resizebox{0.45\textwidth}{!}{\includegraphics[angle=90]
{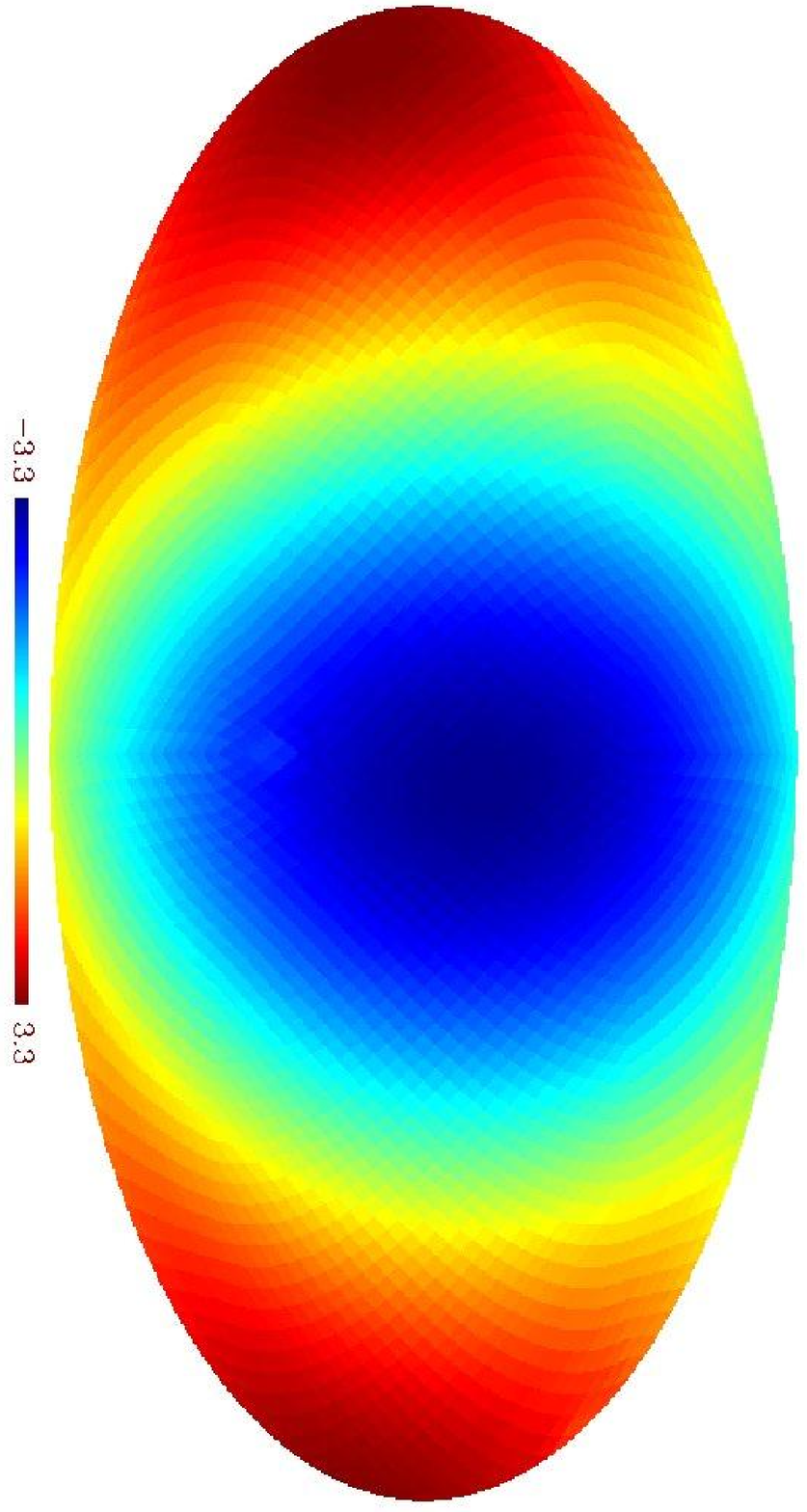}}}
\caption{\label{fig:hitmap} Modulus of the scan tangent vector 
expectation value for {\it WMAP\/} (left), along with the underlying dipole 
field (right, units are mK), in Ecliptic co-ordinates. }
\end{figure}
 
We can introduce a time lag $\delta t$ to characterize the pointing 
offset -- the residual signal at each pointing is then $(T_0/c) \, 
\delta t \,{\bf v} \cdot  \partial {\hat{\bf n}} / \partial t$.  For a 
single beam experiment such as {\em Planck\/} the zeroth order 
approximation for the maximum likelihood solution map is the 
expectation value (in the long-time limit) of the scan tangent vector 
at each pixel (which we define as $\langle \partial {\hat{\bf n}} / 
\partial t \rangle$)  dotted with the dipole velocity. The overall 
amplitude is  proportional to the spin rate of the satellite. 

For an experiment such as {\it WMAP\/} the  $\tilde{\bf t}_0$ map is 
constructed from a difference signal between the two beams, $(T_0/c) \, 
\delta t \,{\bf v} \cdot  \left[ \partial {\hat{\bf n}}_{\rm A} / 
\partial t - \partial {\hat{\bf n}}_{\rm B} / \partial t \right]$ 
(ignoring transmission imbalance factors). The maximum likelihood 
sky-map solution $\tilde{\bf t}$ is also complicated by the fact that  
the $\Sigma$ matrix is not diagonal. We tested the implications of this 
by assuming that {\it WMAP\/} could also operate as a total intensity 
experiment. 

In the default `differencing mode'  with the {\it WMAP\/} scan pattern and 
$\delta t = 25.6\,$ms, we recover the induced quadrupole shown in 
Fig.~\ref{fig:W_solar} to good precision. In `total intensity mode' the 
induced signal was around a factor of several lower. In this case there is a 
subdominant contribution from the  $Y_{4,\,-1}$ harmonic, which has an 
amplitude of around a factor of 4 less than the  $Y_{2,\,-1}$ mode. 
The reason for these results is that, for {\it WMAP}, $ 
\partial {\hat{\bf n}}_{\rm A} / \partial t \simeq  -\partial {\hat{\bf 
n}}_{\rm B} / \partial t$, so the $\tilde{\bf t}_0$ map is similar in 
both differencing and total intensity modes. The $\tilde{\bf t}_0$ map 
contains $Y_{2,\,-1}$ and $Y_{4,\,-1}$ components, but the latter is 
suppressed by the map-making operation in the differencing mode.

The overall structure of the final $\tilde{\bf t}$ maps can therefore 
be understood by considering $\langle \partial {\hat{\bf n}} / \partial 
t \rangle$ and the underlying barycentric dipole field. We show these 
quantities in Ecliptic co-ordinates in  Fig.~\ref{fig:hitmap}, which 
should be compared with the induced map in Fig.~\ref{fig:W_solar}. Near 
the Ecliptic poles pixel scans are approximately isotropic, so the sum 
of dot products with the dipole velocity cancel over the course of a 
year's observations. Along the Plane, the component of the dipole 
velocity parallel to the scan direction is small, since pixels are 
mainly scanned orthogonal to the plane. Moreover, the sum of 
north/south scans cancel over a year. For angles at around
$\theta_{\rm p}$ from 
the poles pixels are mainly scanned in a single direction parallel to 
the equator. Here the dot product with the dipole velocity is large in 
certain azimuthal directions, resulting in a large induced signal. 

We note that it is an unfortunate coincidence that the {\em direction} 
of the dipole field in the Ecliptic Plane results in a quadrupole with 
approximately the same phase as the {\em primordial} signal of {\it WMAP}. 
For the {\it WMAP\/} scan strategy the dipole and quadrupole patterns
therefore have a special and unfortunate relationship.  Although the
pointing offset effect appears not to be real,
one can think of other systematics which could potentially couple to 
the large dipole field in a similar way. These include 
an asymmetric beam  or a time constant in the instrument response. 
Therefore, it seems reasonable that one should take additional care in 
removal of the dipole from the TOD, since the induced signal would be 
similar to the primordial quadrupole.

\begin{figure}
\centering
\mbox{\resizebox{0.45\textwidth}{!}{\includegraphics[angle=90]
{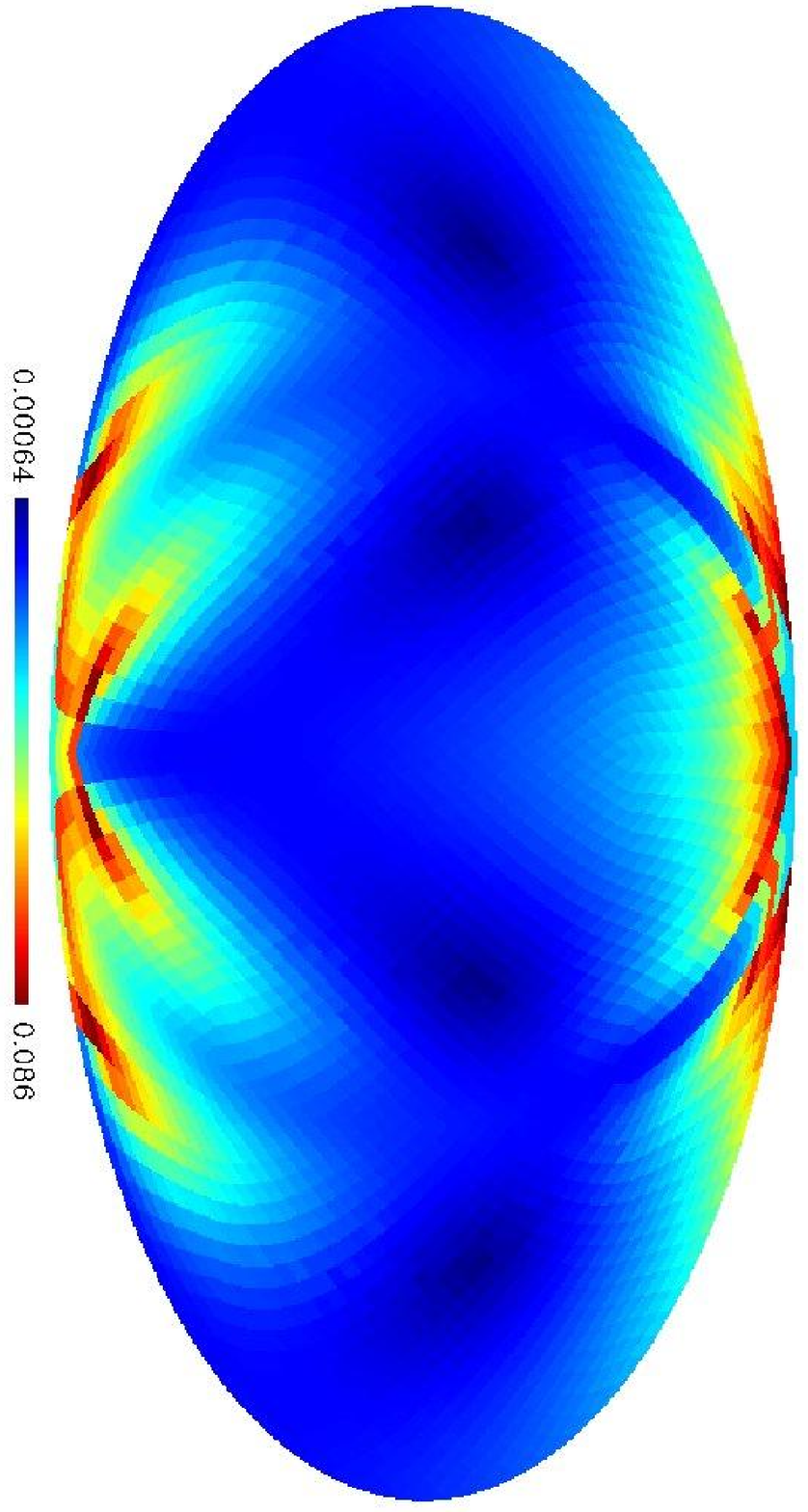}}}
\mbox{\resizebox{0.45\textwidth}{!}{\includegraphics[angle=90]
{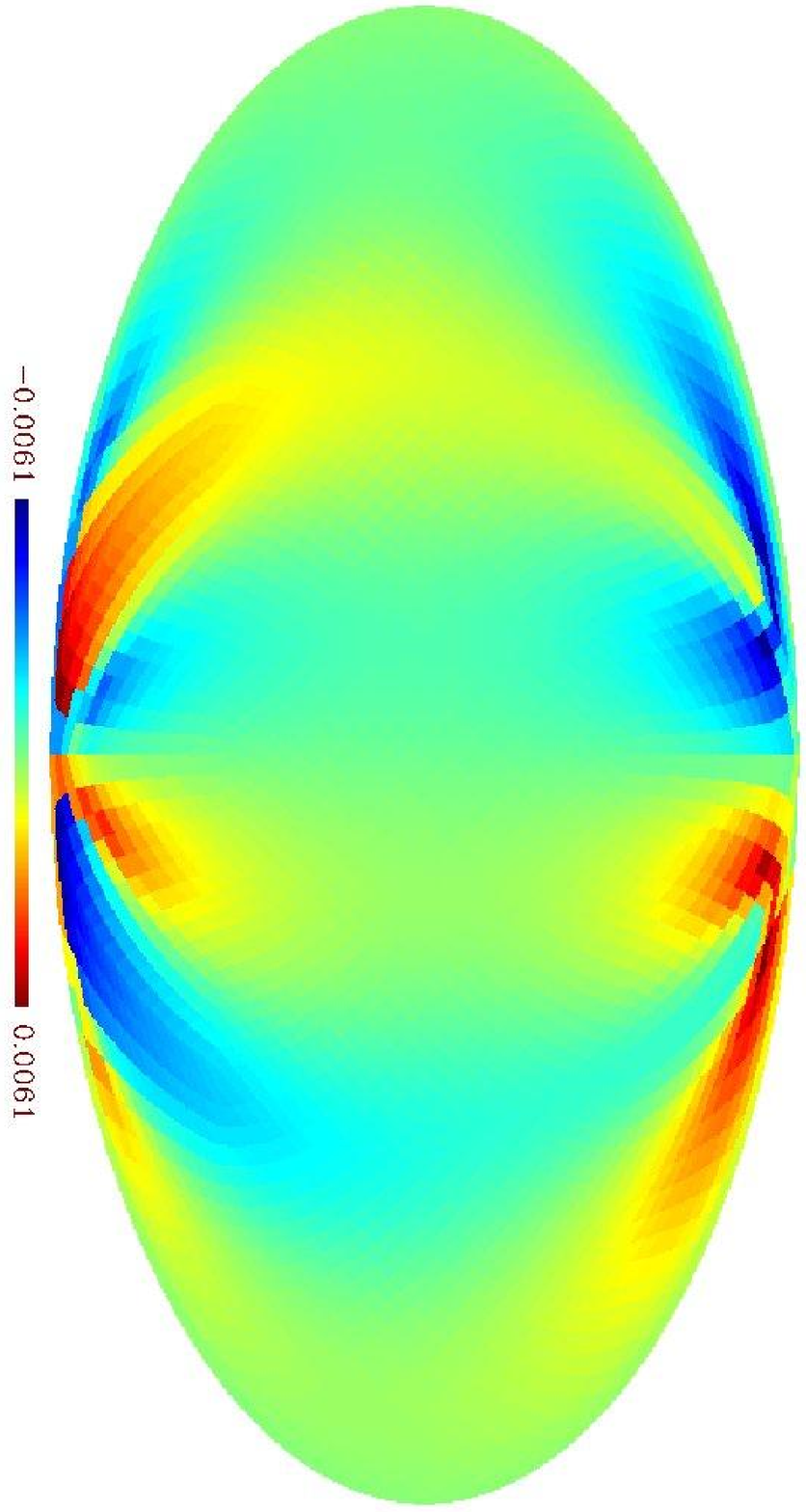}}}
\caption{\label{fig:planck}  Modulus of the scan tangent vector 
expectation value for {\em Planck\/} (left) and the induced map (right, 
units are mK) in Ecliptic co-ordinates. }
\end{figure}

With our toy model in hand,
the scan parameters can be adjusted to test the impact of other scan
strategies on the induced quadrupole map. For the approach of the
{\em Planck\/} satellite the pixel hits map is no longer azimuthally
symmetric (after a year, say) due to the long precession 
period. The modulus of the scan tangent vector, along with the induced 
signal for a time delay of $\delta t = 25.6\,$ms, are shown in 
Fig.~\ref{fig:planck}. The amplitude of the map is similar to
that obtained for {\it WMAP}, but 
the structure is very different due to the different scanning geometry. 
Therefore, one positive aspect of the {\em Planck\/} scan strategy is 
that one would not expect to see similar systematics coupling to the
quadrupole for the {\it Planck\/} scan strategy.

An an aside, our toy model can also be useful in evaluating scan parameters 
of future CMB satellites. For example, the Experimental Probe of Inflationary 
Cosmology ({\it EPIC})~\cite{epic} mission proposal requires that pixel scans 
are as isotropic as possible to minimize polarization angle errors. We 
find that a spin and precession period similar to {\it WMAP}, but with 
$\theta_{\rm b}=45^{\circ}$ and $\theta_{\rm p}=45^{\circ}$, would result in 
improved angular coverage at each pixel compared to {\it WMAP\/}~\cite{gorski}. 

 \section{Observations of point sources}

\begin{figure}
\centering
\mbox{\resizebox{0.3\textwidth}{!}{\includegraphics[angle=0]{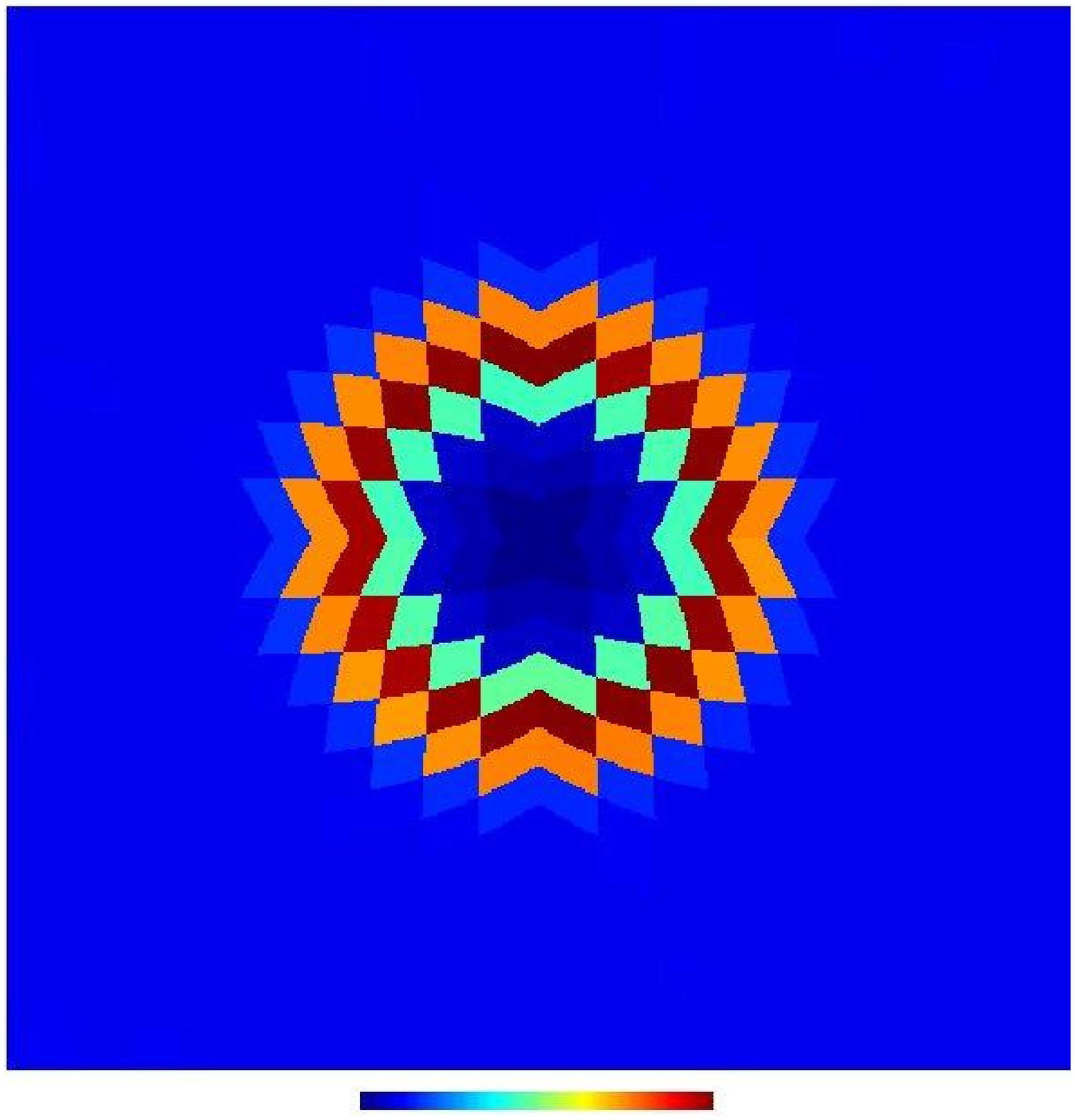}}}
\mbox{\resizebox{0.3\textwidth}{!}{\includegraphics[angle=0]{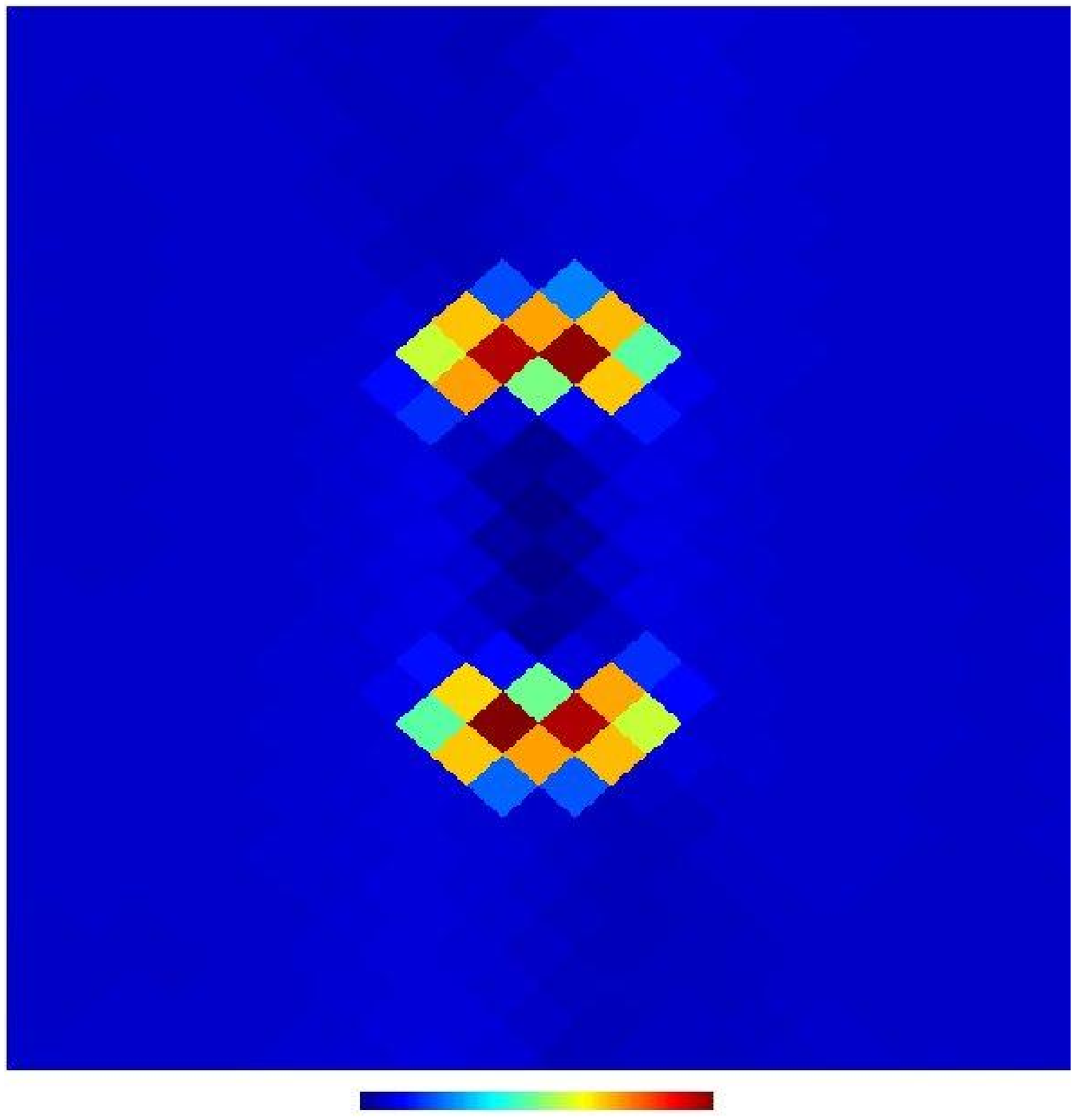}}}
\caption{\label{fig:ps} Observed point source for the {\it WMAP\/} scan 
pattern, originally at an Ecliptic pole (left) and on the equator 
(right). We have set the angular pointing error to be larger than the 
beam size to exaggerate the effect. }
\end{figure}

So far we have focused on the induced signal from the large CMB dipole 
component. Observations of point sources could also leave an 
interesting signature, depending on the instrument beam size compared 
to the angular pointing error $\theta_{\rm e}$. In Fig.~\ref{fig:ps} we 
show an example of two point sources, one originally located at an
Ecliptic pole and one on the equator. We use the {\it WMAP\/} scan strategy and 
set the pointing error to be larger than the beam size to highlight the
effect.  At the pole, 
the source is observed as a circular ring with radius $\theta_{\rm e}$, 
while on the equator there are two arcs in the north/south directions. 
This is consistent with our earlier discussion of the direction of pixel scans 
at these locations.
 
In reality, the beam size of {\it WMAP\/} is larger than the claimed pointing 
offset (for W-band the beam is $\sim 12^\prime)$, so one would observe a 
smearing of sources rather a distinct offset. We expect this effect 
would still be detectable since {\it WMAP\/} has a large catalog of $\sim 400$ 
sources at a signal-to-noise exceeding 5~\cite{Wright:2008ib}. 
Binning sources by Ecliptic latitude and then stacking could lead to 
tight constraints on timing residuals; however, we leave this study for 
future work.

Point sources might also leave additional features in the maps if a
timing residual was present.
When operating in `total intensity 
mode', we find no other features in the map apart from those localized 
around the point source, but in `difference mode' we find 
low amplitude stripes in the map at a distance from the source 
corresponding to twice the beam opening angle. These features are {\em 
only} present when there is a timing residual, and presumedly arise 
from the inability of the maximum likelihood solver to recover a single
solution which is actually the difference between two maps.  One could imagine
in principle adding an additional free parameter to the map-maker to solve
for the timing offset.  However, since the gross effects of the claimed
timing mismatch are not apparent in the data, there seems little point
in investigating the more subtle effects.

Finally, we should mention a related issue, which is that a timing offset
would be at least partially calibrated out in practice.
The {\it WMAP\/} beam boresight is calibrated on twice-yearly 
observations of Jupiter. Therefore, the timing 
offset would be degenerate with the offset of a given detector in the focal 
plane (relative to the boresight) in that {\em particular scan 
direction}. Timing offsets in other scan directions would then only be 
partially corrected by the Jupiter calibration. This could be easily be 
accounted for in our code, however due to uncertainty in the precise 
details of the calibration (e.g.\ Jupiter is observed for 45 day 
intervals twice-yearly), we have not explored this correction in more depth.

\section{Conclusions}

We have investigated recent claims in the literature that the {\it WMAP\/}
quadrupole is systematic in origin, arising
from an offset between the recording of pointing and temperature
data from the satellite. Due to the {\it WMAP\/} scan pattern coupling
with the direction of the dipole field, this effect results in a strong
$Y_{2\,,-1}$ mode in Ecliptic co-ordinates, which happens to be similar
in phase to the actual quadrupole.  We find that
the size of the effect for an offset of $25.6\,$ms 
cannot be large enough to match the observed quadrupole.  We have also
described reasons why the claimed effect is unlikely to be in the
{\it WMAP\/} data, in addition to the {\it WMAP\/} team insisting that any error in their timing  could not be nearly as large as claimed.  Nevertheless, because of
the similarity between the induced and primordial quadrupole signals, one 
should exercise
increased caution in interpreting the amplitude of the primordial 
component.  {\it Planck\/} has a significantly different scan strategy
and will soon provide an independent measurement of the quadrupole.

\section*{Acknowledgments} This research was supported by the Natural 
Sciences and Engineering Research Council of Canada. We thank Kris 
Gorski, Mark Halpern and Hao Liu for useful discussions.

\end{document}